\documentclass[12pt,amsart]{iopart}
\usepackage{fullpage,graphicx,epsf}
\usepackage{iopams}%
\usepackage{amssymb}
\usepackage{amscd}
\usepackage{cite}
\usepackage{color}%
\usepackage{graphicx}
\usepackage{epsfig}
\usepackage[english]{babel}
\usepackage{amsfonts}
\begin{document}

\def\be{\begin{equation}}
\def\ee{\end{equation}}
\def\bea{\begin{eqnarray}}
\def\eea{\end{eqnarray}}
\def\bef{\begin{figure}[h!]}
\def\eef{\end{figure}}
\def\bml{\begin{mathletters}}
\def\eml{\end{mathletters}}
\def\l{\label}
\def\b{\bullet}
\def\no{\nonumber}
\def\fr{\frac}
\def\th{\theta}
\def\eps{\epsilon}
\def\o{\omega}
\def\O{\Omega}
\def\p{\partial}
\title{Relaxation dynamics of stochastic long-range interacting systems}
\author{Shamik Gupta and David Mukamel}
\address{Physics of Complex Systems, Weizmann Institute of Science, Rehovot 76100, Israel}
\ead{shamik.gupta@weizmann.ac.il, david.mukamel@weizmann.ac.il}

\begin{abstract}
Long-range interacting systems, while relaxing towards equilibrium, may get trapped in nonequilibrium quasistationary states (QSS) for a time which diverges algebraically with the system size. These intriguing non-Boltzmann states have been observed under deterministic Hamiltonian evolution of a paradigmatic system, the Hamiltonian Mean-Field (HMF) model. We study here the robustness of QSS with respect to stochastic processes beyond deterministic dynamics within a microcanonical ensemble. To this end, we generalize the HMF model by allowing for stochastic three-particle collision dynamics in addition to the deterministic ones. By analyzing the resulting Boltzmann equation for the phase space density, we demonstrate that in the presence of stochasticity, QSS occur only as a crossover phenomenon over a finite time determined by the strength of the stochastic process. In particular, we argue that the relaxation time to equilibrium does not scale algebraically with the system size. We propose a scaling form for the relaxation time which is in very good agreement with results of extensive numerical simulations. The broader validity of these results is tested on a different stochastic HMF model involving microcanonical Monte Carlo dynamical moves. 
\end{abstract}
\date{\today}
\maketitle

\section{Introduction}
\l{introduction}
In recent years, there has been a surge in interest in systems with long-range interactions. For these systems in $d$ dimensions, the inter-particle potential decays at large separation $r$ as $r^{-\alpha}$, with $\alpha \le d$ (see \cite{Ruffo:2008,Mukamel:2010book, Mukamel:2010} for reviews). Examples are self-gravitating systems \cite{Paddy:1990}, non-neutral plasmas \cite{Nicholson:1992}, dipolar ferroelectrics and ferromagnets \cite{Landau:1960}, two-dimensional geophysical vortices \cite{Chavanis:2002}, wave-particle interacting systems such as a free electron laser \cite{Barre:2004}, and others.

Unlike systems with short-range interactions, long-range interacting systems are non-additive and hence, non-extensive. For example, the total energy of a system of long-range interacting particles homogeneously distributed in a volume $V$ scales superlinearly with the volume as $V^{2-\alpha/d}$. Non-additivity leads to many unusual thermodynamic and dynamical features which are not exhibited by systems with short-range interactions, such as non-concave entropy curves as a function of energy. This implies a negative specific heat at equilibrium within a microcanonical ensemble \cite{Antonov:1962,Lynden-Bell:1968,Thirring:1970,Thirring:1971,Lynden-Bell:1999,Chavanis:2006,Posch:2006}. Since specific heat at canonical equilibrium is always positive, it follows that for long-range systems, the two ensembles need not be equivalent in equilibrium. It has been demonstrated that this inequivalence is particularly manifested whenever the system exhibits a first-order phase transition in the canonical ensemble \cite{Barre:2001,Mukamel:2005}. A recent study of phase transitions and ensemble inequivalence in generic long-range systems has suggested a variety of rich equilibrium behavior \cite{Bouchet:2005}.

As regards dynamical properties, non-additivity often results in breaking of ergodicity in which the phase space is broken up into domains that are not connected by local dynamics \cite{Mukamel:2005,Fel'Dman:1998,Borgonovi:2004,Borgonovi:2006,Hahn:2005,Hahn:2006,Bouchet:2008}. Long-range interactions also lead to violent relaxation \cite{Lynden-Bell:1967} and slow relaxation dynamics whereby a thermodynamically unstable state relaxes to the stable equilibrium state unusually slowly over a timescale which diverges with the system size \cite{Chavanis:2002,Mukamel:2005,Teles:2010}.

A very interesting dynamical feature resulting from long-range interactions is the occurrence of long-lived nonequilibrium quasistationary states (QSS) during relaxation towards equilibrium. These states are characterized by a slow relaxation of macroscopic observables over times which diverge algebraically with the system size. As a result, in the thermodynamic limit, these states do not relax to the Gibbs-Boltzmann equilibrium state so that the system remains trapped in these states in the long time limit.

Much recent exploration of the existence and properties of QSS has been pursued within the ambit of a prototypical model called the Hamiltonian Mean-Field (HMF) model. The model is composed of $N$ classical $XY$ spins which are coupled through mean-field interactions, and is defined by the following Hamiltonian \cite{Ruffo:1995}:
\be
\hspace{-2cm}H=\sum_{i=1}^{N}\fr{p_i^2}{2}+\fr{1}{2N}\sum_{i,j=1}^{N}\left[1-\cos(\th_i-\th_j)\right],
\l{HMF-H}
\ee
where $\th_i \in [-\pi,\pi]$ is the angle of the $i$-th spin, while $p_i \in \mathbb R$ is its angular momentum. The mean-field nature of the interaction makes the model amenable to analytical and numerical studies. As such, it has served as a paradigmatic model to address some characteristic dynamical features typical of long-range interacting systems (for a review on the HMF model, see \cite{Ruffo:2008,Dauxois:2002,Chavanis:2005}). Besides, the model provides a tractable reference to study physical systems like gravitational sheet models \cite{Tsuchiya:1994} and the free-electron laser \cite{Barre:2004}. The magnetic order in the HMF model is characterized by the specific magnetization $\vec{m}$, defined by
\be
\hspace{-2cm}\vec{m}=\left(\fr{1}{N}\sum_{i=1}^{N}\cos \th_i,\fr{1}{N}\sum_{i=1}^{N}\sin \th_i\right)\equiv(m_x,m_y),
\l{m}
\ee
where $m_x$ and $m_y$, respectively, are the $x$ and the $y$ components of the magnetization. The time evolution of the system within a microcanonical ensemble follows the deterministic Hamilton equations of motion, given by
\bea
&&\hspace{-2cm}\fr{d\th_i}{dt}=p_i, \nonumber \\
\l{hameq} \\
&&\hspace{-2cm}\fr{dp_i}{dt}=-m_x\sin \th_i+m_y\cos \th_i. \nonumber
\eea
The dynamics conserves the total energy and momentum. Defining the temperature $T$ as twice the specific kinetic energy, it follows from Eq. (\ref{HMF-H}) that the energy per particle, $\eps$, satisfies the relation $\eps=\fr{T}{2}+\fr{1}{2}(1-m^2)$, where $m^2=m_x^2+m_y^2$. Here and in the rest of the paper, we take the Boltzmann constant to be unity. In equilibrium, the HMF model exhibits a continuous transition from a high-energy paramagnetic phase to a low-energy ferromagnetic one at a critical energy $\eps_c=3/4$, corresponding to a critical temperature $T_c=1/2$. This has been verified within a canonical \cite{Ruffo:1995} as well as a microcanonical ensemble \cite{Barre:2005}. 

In order to study the process of relaxation towards equilibrium under the deterministic dynamics, Eq. (\ref{hameq}), an initial configuration often considered has the angles $\th_i$ independently and uniformly distributed in the interval $[-\pi, \pi]$ and the momenta $p_i$ independently and uniformly distributed in an interval $[-p_0,p_0]$. Here, the parameter $p_0$ fixes the total energy. This state is referred to as the ``water-bag'' initial state. Extensive numerical studies of relaxation have shown that for $\eps$ just below $\eps_c$, the water-bag state is actually quasistationary in the sense that the magnetization fluctuates around its average initial value of zero for a long time which scales algebraically with the system size as $N^\delta$, where $\delta >1$ \cite{Yamaguchi:2003, Yamaguchi:2004, Bouchet:2005rapid}. The final relaxation of magnetization to its non-vanishing equilibrium value occurs over times $t \gg N^\delta$. Several recent studies have revealed that the QSS in the HMF model exhibit many intriguing features like anomalous diffusion \cite{Latora:1999}, non-Gaussian velocity distributions \cite{Latora:2001}, vanishing Lyapunov exponents \cite{Latora:2001}, and others.   

An interesting issue is that of the robustness of QSS with respect to stochastic dynamics beyond deterministic Hamiltonian evolution. Stochastic dynamics may result from the coupling of the system either to an external heat bath or to some internal degrees of freedom. The issue of whether stable QSS emerge under stochastic dynamics has recently been explored for the HMF model evolving within a canonical ensemble \cite{Baldovin:20061,Baldovin:20062,Baldovin:2009}. In these studies, stochasticity is induced into the HMF dynamics through coupling of the system to an external heat bath, with the latter modelled as a short-range interacting system. The coupling allows for energy exchange between the system and the heat bath. As a result, the energy of the system is not conserved by the dynamics. It has been suggested that existence of QSS within a canonical ensemble may depend on the interplay of various timescales in the problem \cite{Baldovin:20061,Baldovin:20062,Baldovin:2009}. It has also been shown that on a timescale that does not diverge with the system size, the system relaxes to the canonical Gibbs-Boltzmann equilibrium state \cite{Baldovin:2009}. Stable QSS were first observed in isolated systems under deterministic Hamiltonian evolution that conserves energy. It is thus of interest to enquire about the stability of QSS in such systems with respect to stochastic dynamics within a microcanonical ensemble, in which the energy of the system is either strictly conserved or fluctuates within a finite energy band. 

In this paper, we study relaxation processes and occurrence of QSS under {\it stochastic dynamics} beyond deterministic Hamiltonian evolution of an isolated system within a {\it microcanonical ensemble}. To this end, we generalize the HMF model to allow for stochasticity in the evolution. Our generalized model follows a piecewise deterministic dynamics in which Hamiltonian evolution, Eq. (\ref{hameq}), is randomly interrupted by stochastic updates of the dynamical variables through three-particle collisions. Namely, three randomly chosen particles collide and their momenta are updated stochastically, while conserving the total energy and momentum and keeping the angular coordinates unchanged. 

To test the broader validity of our study, we also consider a different stochastic process by employing a microcanonical Creutz-like Monte Carlo dynamics \cite{Creutz:1983}. In this dynamics, two randomly chosen particles collide resulting in a stochastic update of their momenta, while conserving the total momentum and keeping the angles unchanged. The update is carried out so long as the energy of the system remains at or below the given initial value. Unlike for three-particle collisions, the system energy is therefore not conserved in collisions, but rather fluctuates below the initial value within a finite energy band which is negligible in the thermodynamic limit.

For the choice of the stochastic dynamics involving three-particle collisions, we study our model by analyzing the Boltzmann equation for the time evolution of the phase space density, by a scaling approach, and by extensive numerical simulations. We propose a scaling form for the relaxation time to equilibrium which is found to be in very good agreement with our results from numerical simulations. We also simulate our model with the stochastic dynamics involving two-particle collisions and find close agreement of the relaxation time with our proposed scaling form. On the basis of our analysis, we suggest that within a microcanonical ensemble and under stochastic dynamics, QSS occur only as a crossover phenomenon over a characteristic time which is determined by the strength of the stochastic process. In particular, unlike the purely deterministic case, the relaxation time at long times does not scale algebraically with the system size. The crossover timescale diverges when the rate for interparticle collision vanishes; only then quasistationarity is restored. A brief account of some of the results obtained in the present study is given in \cite{Gupta:2010letter}, and the purpose here is to present a detailed derivation of these results as well as to report on additional findings.  

The paper is organized as follows. In Section \ref{3collisions}, we define the generalized HMF model with the stochastic process of three-particle collisions. We recall how QSS emerge under deterministic dynamics, and then analyze in detail the occurrence of QSS in our model by considering the Boltzmann equation of our model. We demonstrate that under stochastic evolution, QSS occur only as a crossover phenomenon. A scaling form for the relaxation time to equilibrium is proposed which is verified by extensive numerical simulations. In Section \ref{2collisions}, we investigate the broader validity of our results by considering our generalized model with a different stochastic process, namely, microcanonical Monte Carlo process involving two-particle collisions. We find numerical agreement of the relaxation time with our proposed scaling form. The paper ends with concluding remarks in Section \ref{conclusions}.   
\section{Generalized HMF model with three-particle collisions}
\l{3collisions}
In this section, we introduce our generalized HMF model with the stochastic process of three-particle collisions discussed in Section \ref{introduction}. We then go on to investigate in detail the relaxation processes and existence of QSS in this model.
\subsection{The model}
\l{themodel}
Our model evolves by the following repetitive sequence of events. A deterministic evolution according to Eq. (\ref{hameq}) occurs for a random time interval $\tau$, distributed as $\alpha e^{-\alpha \tau}$. This is followed by an instantaneous sweep consisting of $N^3$ collisions. Thus, the dynamics of the model is piecewise deterministic for uncorrelated random intervals of time. Now, we specify the process of collision. In each collision, three particles $(i,j,k)$ are randomly chosen and their momenta are updated stochastically, $(p_i,p_j,p_k) \rightarrow (q_i,q_j,q_k)$, while conserving the total energy and momentum and keeping the angles $(\th_i,\th_j,\th_k)$ unchanged. Note that the parameter $\alpha$ has the dimension of 1/time and sets the timescale for collisions: on average, each triplet of particles undergoes one collision after every time interval $\alpha^{-1}$. 

We now consider the Boltzmann equation of our model which governs the time evolution of the single-particle phase space distribution $f(\th,p,t)$ for infinite $N$. The quantity $f(\th,p,t)d\th dp$ gives the probability that at time $t$, a randomly chosen particle has its angle between $\th$ and $\th+d\th$ and its momentum between $p$ and $p+dp$. The distribution is normalized to unity, $\int d\th ~dp ~f(\th,p,t)=1$, and is also periodic in $\th$, $f(\th+2\pi,p,t)=f(\th,p,t)$, at all times. The Boltzmann equation of our model is given by
\be
\hspace{-2cm}\fr{\p f}{\p t}+p\fr{\p f}{\p \th}-\fr{\p \langle v \rangle}{\p \th}\fr{\p f}{\p p}=\left(\fr{\p f}{\p t}\right)_c,\l{Boltz}
\ee
where
\be
\hspace{-2cm}\left(\fr{\p f}{\p t}\right)_c=\int d\eta ~R\left[f(\th,q,t)f(\th',q',t)f(\th'',q'',t)-f(\th,p,t)f(\th',p',t)f(\th'',p'',t)\right], \l{collisionterm}
\ee
\be
\hspace{-2cm}R=\alpha\delta(p+p'+p''-q-q'-q'')\delta\left(\fr{1}{2}(p^2+p'^2+p''^2)-\fr{1}{2}(q^2+q'^2+q''^2)\right),\l{Rdefinition}
\ee
and $d\eta \equiv dp'dp''dqdq'dq''d\th'd\th''$. In Eq. (\ref{Boltz}), the average potential $\langle v\rangle$ is given by
\be
\hspace{-2cm}\langle v\rangle=\int dp' ~d\th'~\left[1-\cos(\th-\th')\right]f(\th',p',t).
\ee
Equation (\ref{collisionterm}) represents the three-body collision term. $R$ is the rate for collisions $(p,p',p'') \rightarrow (q,q',q'')$ that conserve energy and momentum and keep the angles $(\th,\th',\th'')$ of the three colliding particles unchanged.

The Boltzmann equation with a similar three-particle collision term in one dimension has been considered earlier \cite{Ma:1983}. In that work, the particle coordinates $x$ ($\equiv \th$ in our notation) refer to the physical space. As a result, a collision takes place only when the coordinates of the colliding particles are the same. By contrast, the particle coordinates $\th$ in our model represent internal degrees of freedom in the $XY$ ``spin'' space, so that they need not be equal for the colliding particles.  

In the absence of collisions ($\alpha=0$), our generalized model reduces to the deterministic HMF model discussed in the Introduction. We refer to the Boltzmann equation with $\alpha=0$ as the Vlasov-equation limit \cite{Nicholson:1992}. Note that both the Boltzmann and the Vlasov equations are valid in an infinite system. For finite $N$, both these equations have size-dependent correction terms, and will thus be valid only for times when these corrections may be neglected.

Let us remark on some general characteristics of the stationary solutions of Eq. (\ref{Boltz}). These will have bearings on our subsequent discussions on the existence of QSS in our generalized model. In the Vlasov limit, any state which is homogeneous in $\th$ but with an arbitrary normalized momentum distribution $f_0(p)$ represents a stationary solution: $f^\mathrm{st}(\th,p)=g(\th)f_0(p)$, where the label ``st" stands for ``stationary". Here, $g(\th)=1/2\pi$ for $\th \in [-\pi,\pi]$ and is zero otherwise. By contrast, of all possible states with homogeneous $\th$, only the one with a Gaussian distribution of the momentum is stationary under the Boltzmann equation: $f^\mathrm{st}(\th,p)=g(\th)\fr{1}{\sqrt{2\pi T}}e^{-p^2/2T}$. This may be easily seen by direct substitution into Eq. (\ref{Boltz}). Since the average magnetization in a homogeneous state is zero, the temperature $T$ is related to the energy density through $T=2\eps-1$.

With the above background, we first recollect known results on how QSS emerge in the Vlasov limit. We next address the issue of QSS in our generalized model.
\subsection{QSS under deterministic dynamics: A short recapitulation}
\l{deterministic}
As discussed above, in an infinite system and with deterministic dynamics ($\alpha=0$), the phase space evolution follows the Vlasov equation. This equation admits a stationary state which is homogeneous in $\th$, but has an arbitrary momentum distribution. Here, QSS are related to stable stationary solutions of the Vlasov equation. For example, consider an initial state which is homogeneous in angles and uniform in momenta (the ``water-bag'' initial condition). This state may be prepared by sampling independently for each of the $N$ spins the angle $\th_i$ uniformly in $[-\pi,\pi]$ and the momentum $p_i$ uniformly in $[-p_0,p_0]$. The corresponding initial phase space distribution is given by
\be
f(\th,p,0)=\left \{ \begin{array}{ll}
                      \fr{1}{2\pi}\fr{1}{2p_0}, & \mbox{for $\th \in [-\pi,\pi]$ and $p \in [-p_0,p_0]$} \\
                      0, & \mbox{otherwise.}
                      \end{array}
                      \right.
\l{water-bag}
\ee
The average magnetization of this state is zero with fluctuations of order $1/\sqrt{N}$, while the energy density is $\eps=p_0^2/6+1/2$. 

Earlier studies have found that the water-bag initial state is linearly stable under the Vlasov equation in the energy range $\eps^*=7/12 < \eps <\eps_c=3/4$ and is unstable below $\eps^*$ \cite{Yamaguchi:2004, Jain:2007, Chavanis:2009}. In the stable regime, the magnetization stays close to its average initial value of zero and evolves only over times when finite-size corrections in the Vlasov equation become appreciable. The corresponding characteristic timescale has been found to grow with the system size as $N^\delta$, where $\delta>1$ \cite{Bouchet:2005rapid}. For instance, numerical studies at $\epsilon=0.69$ give $\delta \simeq 1.7$ \cite{Yamaguchi:2003, Yamaguchi:2004}. Hence, for $\eps^* < \eps < \eps_c$, the water-bag initial state is quasistationary. The final relaxation of magnetization to equilibrium at times $t \gg N^\delta$ may be represented as
\be
\hspace{-2cm}m(t) \sim \fr{1}{\sqrt{N}}e^{\fr{t}{N^{\delta}}} \mathrm{~~~~for~~~} t \gg N^{\delta},
\l{mstable}
\ee
where the prefactor accounts for fluctuations in the initial state. It follows from Eq. (\ref{mstable}) that the relaxation time $\tau(N)$, namely the time taken for the magnetization to reach the equilibrium value of $O(1)$, diverges as $N^{\delta}\ln N$. On the other hand, for energies below $\eps^*$, the water-bag distribution is linearly unstable and the initial magnetization relaxes exponentially fast \cite{Jain:2007}: 
\be
\hspace{-2cm}m(t) \sim \fr{1}{\sqrt{N}}e^{\Gamma t} \mathrm{~~~~for~~~} t \gg \fr{1}{\Gamma},
\l{munstable}
\ee
where $\Gamma^2=6\left(\fr{7}{12}-\epsilon\right)$ is independent of $N$. Therefore, there are no QSS for $\eps < \eps^*$. In this regime, the relaxation time $\tau(N)$ scales as $\ln N$. 

Let us also mention here the relaxation of a homogeneous initial state with another typical momentum distribution, namely, a Gaussian distribution. This initial state is linearly unstable at all energies below $\eps_c$. As a result, relaxation occurs exponentially fast over a timescale $\tau(N) \sim \ln N$ and there are no QSS \cite{Yamaguchi:2004, Jain:2007}.    

\subsection{QSS under stochastic dynamics}
\l{stochastic}
Here, we analyze in detail the existence of QSS in our generalized HMF model while starting from a water-bag state. We begin by presenting some simple considerations for the behavior of this initial state under the Boltzmann equation of our model. We then analyze the Boltzmann equation, study the behavior of magnetization, first in an infinite system and then in a finite one. The analysis in a finite system is performed by employing a scaling approach. We finally test our predictions by numerical studies.

\subsubsection{Simple considerations}
\l{simpleconsiderations}
As discussed in Section \ref{themodel}, the phase space evolution of our model in the limit of infinite $N$ follows the Boltzmann equation, and the only stationary state which is homogeneous in $\th$ has Gaussian-distributed momenta. It therefore follows that this state has an associated fixed point in the space of different possible states with homogeneous $\th$ and arbitrary momentum distribution which will all evolve under the dynamics towards this state. We call this fixed point the Gaussian fixed point. Note that the stationary Gibbs-Boltzmann equilibrium state also has Gaussian-distributed momenta, but has non-homogeneous $\th$ distribution at energies below $\eps_c$. 

A water-bag initial state is not stationary under the Boltzmann equation. Consequently, the initial uniform momentum distribution will evolve under the dynamics towards the stationary Gaussian momentum distribution. Interestingly, while the momentum distribution evolves, the $\th$ distribution (and consequently, the magnetization) does not change in time. This is easily seen by substituting the water-bag initial state, Eq. (\ref{water-bag}), into the Boltzmann equation (\ref{Boltz}). The water-bag state will thus evolve under the dynamics towards the state associated with the Gaussian fixed point over the timescale $\alpha^{-1}$, the only timescale in the problem. 

Anticipating that a dynamical flow may exist from the Gaussian fixed point towards the equilibrium Gibbs-Boltzmann fixed point at energies $\eps < \eps_c$, one may ask whether the former is dynamically stable with respect to perturbations in both the momentum and the $\th$ distributions. In the Vlasov limit (deterministic dynamics with no collisions), the Gaussian fixed point is linearly unstable under such a perturbation at all energies below $\eps_c$ \cite{Yamaguchi:2004, Jain:2007}. In the following section, we explore its stability under the Boltzmann equation and demonstrate that in this case too, it is linearly unstable below the critical point. As we will show later, this instability results in a fast relaxation towards the equilibrium Gibbs-Boltzmann state in our generalized model. 

\subsubsection{Analysis of the Boltzmann equation}
\l{Boltzmannanalysis}
In this section, we prove that the homogeneous state with Gaussian-distributed momenta is linearly unstable under the Boltzmann equation for energies $\eps < \eps_c$. Our proof is a generalization of that performed in the Vlasov limit in Ref. \cite{Jain:2007}. In particular, we demonstrate the instability in the limit of small $\alpha$.

The stability analysis is carried out by linearizing Eq. (\ref{Boltz}) about the homogeneous state. To this end, assuming $\th \in [-\pi,\pi]$, we write
\be
\hspace{-2cm}f(\th,p,t)=\fr{1}{2\pi}f_0(p)\left[1+\lambda \phi(\th,p,t)\right],
\l{expansion}
\ee
where 
\be
\hspace{-2cm}f_0(p)=\fr{e^{-p^2/2T}}{\sqrt{2\pi T}}; ~~~~~~~~p \in [-\infty, \infty]~~\mathrm{and}~~T=2\eps-1.
\l{fp}
\ee
Since the initial angles and momenta of the $N$ spins are sampled independently according to the distribution $\fr{1}{2\pi}f_0(p)$, the small parameter $\lambda$ in Eq. (\ref{expansion}) is of order $1/\sqrt{N}$. We next substitute Eq. (\ref{expansion}) into Eq. (\ref{Boltz}), keep only first-order terms in $\lambda$, and finally use the equality $f_0(q)f_0(q')f_0(q'')=f_0(p)f_0(p')f_0(p'')$ which follows from the energy conservation condition demanded by $R$. We get 
\bea
&&\hspace{-2cm}\fr{\p \phi(\th,p,t)}{\p t}+p\fr{\p \phi(\th,p,t)}{\p \th}-\fr{1}{2\pi f_0(p)}\fr{\p f_0(p)}{\p p}\int dp' ~d\th'~\sin(\th-\th')f_0(p')\phi(\th',p',t) \nonumber \\
&&\hspace{-2cm}=\fr{1}{(2\pi)^2}\int d\eta ~f_0(p')f_0(p'')R[\phi(\th,q,t)+\phi(\th',q',t)+\phi(\th'',q'',t)\nonumber \\
&&-\phi(\th,p,t)-\phi(\th',p',t)-\phi(\th'',p'',t)].
\l{linearizedeqn}
\eea

Any arbitrary $\phi(\th,p,t)$ cannot be expanded in terms of the eigenmodes of the linearized equation (\ref{linearizedeqn}) \cite{Nicholson:1992}. However, when unstable modes exist, the dynamics at sufficiently long times is dominated by the largest of the eigenmodes. Let $\o$ denote the corresponding frequency. Since $\phi(\th,p,t)$ is periodic in $\th$, it may be expanded in a Fourier series. We finally have
\be
\hspace{-2cm}\phi(\th,p,t)=\sum_k \phi_k(p,\o)e^{i(k\th+\o t)}.
\l{Fourier}
\ee
We note that the last term on the left hand side of Eq. (\ref{linearizedeqn}), which represents coupling between the spins, involves $e^{\pm i\th}$. As a result, only the modes with $k=\pm 1$ are affected by the inter-particle interaction potential and are therefore relevant for our studies. After substituting Eq. (\ref{Fourier}) into Eq. (\ref{linearizedeqn}), we find that the coefficients $\phi_{\pm 1}$ satisfy the following equation.
\bea
&&\hspace{-2cm}i(\o\pm p)\phi_{\pm 1}(p,\o)\mp\fr{1}{f_0(p)}\fr{\p f_0(p)}{\p p}\int \fr{dp'}{2i}~f_0(p')\phi_{\pm 1}(p',\o)\nonumber \\
&&\hspace{-2cm}=\fr{1}{(2\pi)^2}\int d\eta ~f_0(p')f_0(p'')R[\phi_{\pm 1}(q,\o)-\phi_{\pm 1}(p,\o)].
\l{eigenvalueeqn}
\eea
In obtaining the above equation from Eq. (\ref{linearizedeqn}), we have used the fact that the terms involving $\phi(\th',q',t), \phi(\th'',q'',t), \phi(\th',p',t),$ and $\phi(\th'',p'',t)$ in the latter equation do not contribute to the modes with $k=\pm 1$.

Treating the term on the right hand side of Eq. (\ref{eigenvalueeqn}) as a small perturbation in the limit $\alpha \rightarrow 0$, we now solve the above equation for the eigenfrequency $\o$ to lowest order in $\alpha$. We begin by discussing the unperturbed solution, $\phi^{0}_{\pm 1}(p,\o_0)$, corresponding to the Vlasov limit ($ \alpha=0$) of the Boltzmann equation. It satisfies 
\be
\hspace{-2cm}i(\o_0 \pm p)\phi^{0}_{\pm 1}(p,\o_0)\mp \fr{1}{f_0(p)}\fr{\p f_0(p)}{\p p}\int \fr{dp'}{2i}~f_0(p')\phi^{0}_{\pm 1}(p',\o_0)=0.
\l{unperturbedeigenvalueeqn}
\ee
Using $\fr{\p f_0(p)}{\p p}=-\fr{p}{T}f_0(p)$, Eq. (\ref{unperturbedeigenvalueeqn}) yields
\be
\hspace{-2cm}\phi^{0}_{\pm 1}(p,\o_0)=\fr{pI_\pm}{2T(p \pm \o_0)},
\l{unperturbedfn}
\ee
where $I_\pm$ is given by 
\be
\hspace{-2cm}I_\pm=\int_{-\infty}^\infty dp ~f_0(p)\phi^{0}_{\pm 1}(p,\o_0).
\l{Ipmdefn}
\ee
To determine the unperturbed eigenfrequency $\o_0$, multiply both sides of Eq. (\ref{unperturbedfn}) by $f_0(p)$ given in Eq. (\ref{fp}), and then integrate over $p$. We get 
\be
\hspace{-2cm}I_\pm (1-J_\pm)=0,
\l{unperturbedomegacondn}
\ee
where 
\be
\hspace{-2cm}J_\pm = \fr{1}{2T\sqrt{2\pi T}}\int_{-\infty}^{\infty}dp ~\fr{pe^{-p^2/2T}}{p \pm \o_0}=\fr{1}{T\sqrt{2\pi T}}\int_{0}^{\infty}dp ~\fr{p^2e^{-p^2/2T}}{p^2 - \o_0^2}.
\ee
From Eq. (\ref{unperturbedomegacondn}), since $I_\pm \ne 0$, the condition determining the frequency $\o_0$ is $J_\pm = 1$, i.e., 
\be
\hspace{-2cm}\fr{1}{T\sqrt{2\pi T}}\int_{0}^{\infty}dp ~\fr{p^2e^{-p^2/2T}}{p^2 - \o_0^2}=1.
\l{unperturbedomegacondnfinal}
\ee 

In the unperturbed case, it is known that the homogeneous state with Gaussian-distributed momenta is linearly unstable at all energies below the critical point \cite{Jain:2007}. The corresponding frequencies are obtained by evaluating the integral in Eq. (\ref{unperturbedomegacondnfinal}) for $\o_0^2=-\O_0^2$. Here, real $\O_0 \ge 0$ implies instability of the corresponding Fourier mode. The result is \cite{Jain:2007}
\be
\hspace{-2cm}\fr{T_c}{T}-\fr{\sqrt{\pi}}{(2T)^{3/2}}\O_0 e^{\O_0^2/2T}\mathrm{Erfc}\left[\fr{\O_0}{\sqrt{2T}}\right]=1,
\l{unperturbedomega}
\ee
where $\mathrm{Erfc}[x]=\fr{2}{\sqrt{\pi}}\int_x^\infty dt ~e^{-t^2}$ is the complementary error function. From Eq. (\ref{unperturbedomega}), it follows that the point of neutral stability ($\O_0=0$) coincides with the critical point $T=T_c=1/2$ \cite{Yamaguchi:2004, Jain:2007}. Just below $\eps_c$, Eq. (\ref{unperturbedomega}) gives, to leading order in $T_c-T$,
\be
\hspace{-2cm}\O_0 \approx \fr{2}{\sqrt{\pi}}(T_c-T)=\fr{4}{\sqrt{\pi}}(\eps_c-\eps).
\l{unperturbedomegasmall}
\ee

We now proceed to obtain the perturbed eigenfrequency to lowest order in $\alpha$. On substituting the unperturbed solutions into Eq. (\ref{eigenvalueeqn}) and using Eq. (\ref{unperturbedeigenvalueeqn}), we find that, to $O(\alpha)$, the change in the eigenfrequency $\o_0$, namely $\Delta \o\equiv (\o-\o_0)$, satisfies  
\be
\hspace{-2cm}i\Delta \o ~\phi^{0}_{\pm 1}(p,\o_0)=\fr{1}{(2\pi)^2}\int d\eta ~f_0(p')f_0(p'')R[\phi^{0}_{\pm 1}(q,\o_0)-\phi^{0}_{\pm 1}(p,\o_0)].
\l{perturbedeigenvalueeqn}
\ee
Multiplying both sides of this equation by $f_0(p)$ and then integrating over $p$, we get
\be
\hspace{-2cm}i\Delta \o I_\pm=\fr{1}{(2\pi)^2}\int dpdp'dp''dqdq'dq''d\th'd\th'' ~f_0(p)f_0(p')f_0(p'')R[\phi^{0}_{\pm 1}(q,\o_0)-\phi^{0}_{\pm 1}(p,\o_0)],
\ee
where we have explicitly written down the form of $d\eta$ in Eq. (\ref{perturbedeigenvalueeqn}). Performing the integration over $\th'$ and $\th''$, we finally get
\be
\hspace{-2cm}i\Delta \o I_\pm=I-\int_{-\infty}^\infty dp ~f_0(p)\phi^{0}_{\pm 1}(p,\o_0)\nu(p),
\l{perturbedeigenvalueeqnsimple}
\ee
where 
\be
\hspace{-2cm}I=\int dp dp'dp''dqdq'dq'' ~f_0(p)f_0(p')f_0(p'')R\phi^{0}_{\pm 1}(q,\o_0),
\l{Idefinition}
\ee
and
\be
\hspace{-2cm}\nu(p)=\int dp'dp''dqdq'dq'' ~f_0(p')f_0(p'')R.
\l{nudefinition}
\ee

We next evaluate the integrals in Eqs. (\ref{Idefinition}) and (\ref{nudefinition}). To do this, we transform to a new set of variables, as discussed in Ref. \cite{Ma:1983}. Let $P$ denote the three-particle momentum, given by
\be
\hspace{-2cm}P=p+p'+p'',
\l{Pdefinition}
\ee
and let $E$ denote the three-particle energy, given by
\be
\hspace{-2cm}E=\fr{1}{2}(p^2+p'^2+p''^2).
\l{Edefinition}
\ee
In the collision process $(p,p',p'') \rightarrow (q,q',q'')$, both $P$ and $E$ are conserved. As a result, the updated momenta, $(q,q',q'')$, lie on a circle formed by the intersection of the plane given by Eq. (\ref{Pdefinition}) and the spherical surface given by Eq. (\ref{Edefinition}). The radius of this circle is $r=\sqrt{2E-P^2/3}$. Note that $P \in [-\infty,\infty]$, while $r \in [0, \infty]$. One may parametrize the new momenta, $(q,q',q'')$, in terms of an angle $\psi \in [0,2\pi]$ measured along the circle of intersection and write \cite{Ma:1983}
\bea
&&\hspace{-2cm}q=\fr{P}{\sqrt{3}}+\sqrt{\fr{2}{3}}r\cos \psi,\l{qdefinition}\\
&&\hspace{-2cm}q'=\fr{P}{\sqrt{3}}-\fr{r}{\sqrt{6}}\cos \psi-\fr{r}{\sqrt{2}}\sin \psi, \l{q'definition} \\ 
&&\hspace{-2cm}q''=\fr{P}{\sqrt{3}}-\fr{r}{\sqrt{6}}\cos \psi+\fr{r}{\sqrt{2}}\sin \psi.\l{q''definition}
\eea
Then, one has \cite{Ma:1983}
\be
\hspace{-2cm}dqdq'dq''R=\fr{\alpha}{\sqrt{3}}d\psi.
\l{integrationmeasuredefinition1}
\ee
Using Eq. (\ref{integrationmeasuredefinition1}) and Eq. (\ref{fp}) in Eq. (\ref{nudefinition}), one finds that $\nu(p)$ evaluates to a constant:
\be
\hspace{-2cm}\nu(p)=\fr{\alpha 2\pi}{\sqrt{3}}.
\l{nuvalue}
\ee
The integral $I$ in Eq. (\ref{Idefinition}) is evaluated in \ref{appendix1} to yield
\be
\hspace{-2cm}I=\fr{2\pi \alpha T_c I_\pm}{T\sqrt{3}}-\fr{\pi^{3/2}\alpha I_\pm}{T^{3/2}\sqrt{30}}\left[\O_0 +O(\O_0^2)\right].
\l{Ifinalresultfromappendix}
\ee

On using Eqs. (\ref{nuvalue}), (\ref{Ifinalresultfromappendix}) and (\ref{Ipmdefn}) in Eq. (\ref{perturbedeigenvalueeqnsimple}), we finally  get
\be
\hspace{-2cm}\O=\O_0+\fr{2\pi\alpha (T_c-T)}{T \sqrt{3}}-\fr{\pi^{3/2}\alpha}{T^{3/2}\sqrt{30}}\left[\O_0+O(\O_0^2)\right],
\l{omegamid}
\ee
where we have substituted $\o_0=-i\O_0$ in Eq. (\ref{perturbedeigenvalueeqnsimple}) to get the perturbed frequency $\O=i\o$. From Eq. (\ref{unperturbedomega}), we have
\be
\hspace{-2cm}\fr{T_c-T}{T}=\fr{\sqrt{\pi}}{(2T)^{3/2}}\O_0 e^{\O_0^2/2T}\mathrm{Erfc}\left[\fr{\O_0}{\sqrt{2T}}\right],
\ee
which, on using in Eq. (\ref{omegamid}), gives
\be
\hspace{-2cm}\O=\O_0\left[1+\alpha \left\{A'+O(\O_0)\right\}\right]; ~~~~~~~~~~A'=\fr{\pi^{3/2}}{\sqrt{6}T^{3/2}}\left(1-\fr{1}{\sqrt{5}}\right).
\l{omegafullexpression}
\ee
Equation (\ref{omegafullexpression}) shows that the leading behavior of $\O$ is determined by the unperturbed frequency $\O_0$. This implies that in the presence of collisions, the homogeneous state with Gaussian-distributed momenta is unstable at all energies below $\eps_c$. Just below $\eps_c$, Eq. (\ref{omegafullexpression}) gives 
\be
\hspace{-2cm}\O \approx \O_0[1+\alpha A]; ~~~~~~~~~~A=\fr{2\pi^{3/2}}{\sqrt{3}}\left(1-\fr{1}{\sqrt{5}}\right),
\ee
with $\O_0$ given in Eq. (\ref{unperturbedomegasmall}).

Having demonstrated linear instability of a homogeneous state with Gaussian-distributed momenta under the Boltzmann equation, we now proceed to discuss the evolution of magnetization in our generalized model, first in an infinite system, and then, in a finite one. For the latter, we invoke a scaling approach to discuss the relaxation to equilibrium. 

\subsubsection{Behavior of magnetization in an infinite system}
\l{magnetizationinfinitesystem}
As mentioned in Section \ref{simpleconsiderations}, the water-bag initial state evolves towards the state associated with the Gaussian fixed point (G) over the timescale $\alpha^{-1}$. Based on our analysis in Section \ref{Boltzmannanalysis}, we expect that the instability of the fixed point G with respect to perturbations in both the $\th$ and the $p$ distributions generates a dynamical flow to the equilibrium Gibbs-Boltzmann fixed point (GB), which has Gaussian-distributed $p$ and non-homogeneous $\th$. This is shown schematically in Fig. \ref{flowdiagram}. Note that the $\th$-distribution is non-homogeneous only below $\eps_c$.

\bef
\begin{center}
\includegraphics[width=100mm]{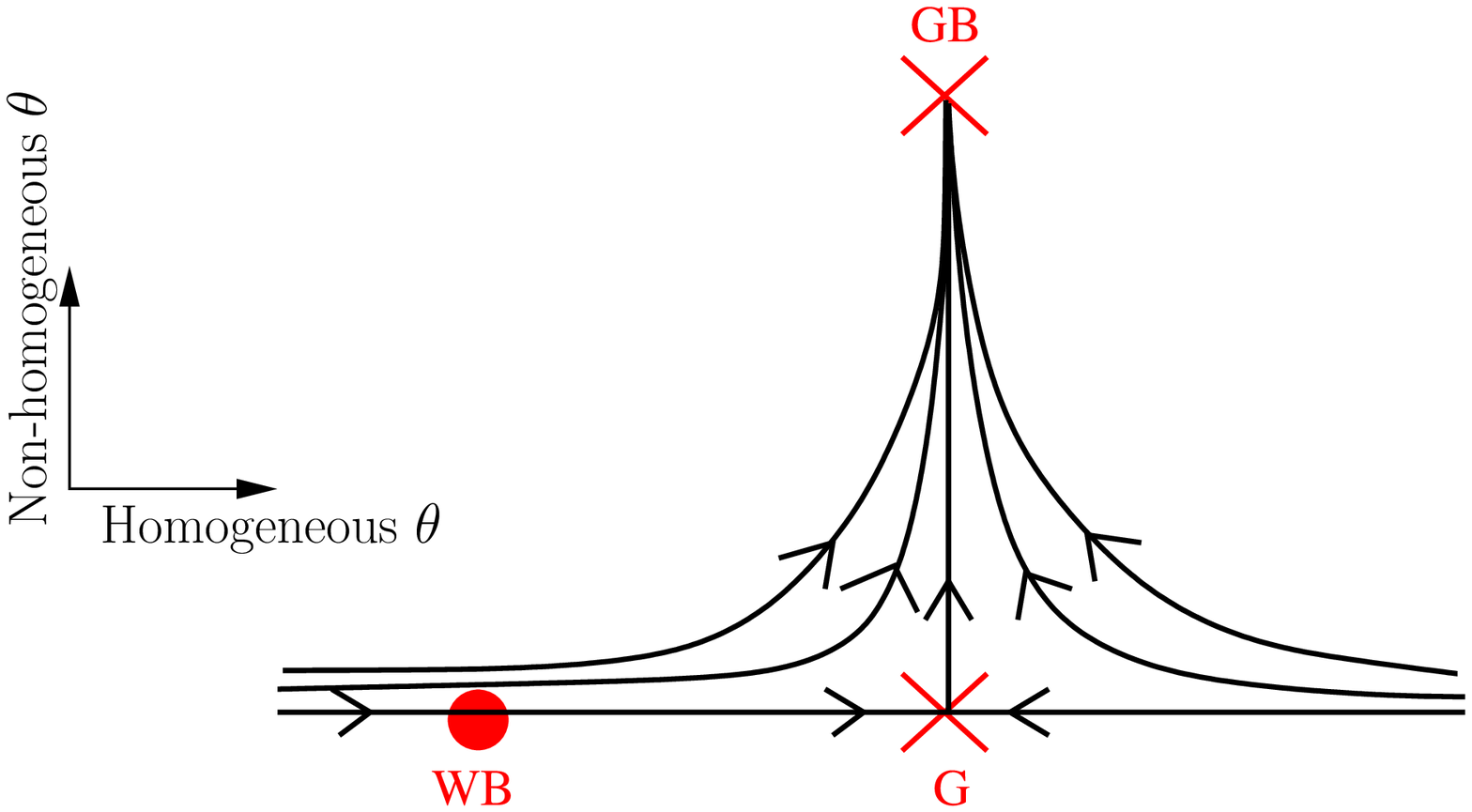}
\vspace{-5mm}
\caption{Schematic representation of relative stabilities of the Gaussian fixed point (G) and the Gibbs-Boltzmann fixed point (GB) under evolution governed by the Boltzmann equation for energies $\eps < \eps_c$. The fixed points are denoted by crosses. The water-bag state, WB, does not have an associated fixed point, and is therefore marked by a filled circle. The horizontal axis represents states which are homogeneous in $\th$, while the vertical axis represents those with non-homogeneous $\th$.}
\l{flowdiagram}
\end{center}
\eef

On starting from a state with almost homogeneous $\th$ distribution and an arbitrary $p$ distribution, the drift towards the Gaussian fixed point G will cause a dynamical flow along that direction over the timescale $\alpha^{-1}$. However, since G is unstable to inhomogeneous $\th$ distribution, the system never gets close to it and it eventually relaxes towards the stable Gibbs-Boltzmann fixed point GB. This is indicated schematically in Fig. \ref{flowdiagram} by curved trajectories starting close to the horizontal axis for homogeneous $\th$ distribution and ending in GB. Therefore, for non-vanishing $\alpha$ and in an infinite system, we expect an initial state close to a water-bag state to relax towards equilibrium over the timescale $\alpha^{-1}$. Armed with this background, we discuss in the next section the relaxation dynamics of the water-bag initial state in a finite system by employing a scaling approach.

\subsubsection{Behavior of magnetization in a finite system: A scaling approach}
\l{magnetizationfinitesystem}

To discuss the relaxation of the magnetization from a water-bag initial state in our generalized model in a finite system, we first note that there are two relevant timescales in the problem, (i) the timescale $\sim \alpha^{-1}$ over which collisions occur in the system, and (ii) the timescale $\sim N^\delta$ over which size-dependent correction terms in the Vlasov equation become appreciable in governing the dynamics. It is thus natural to invoke a scaling approach to analyze the interplay between the two timescales.

Consider the limit $\alpha^{-1} \ll N^\delta$ and $\alpha^{-1} \ll t \ll N^\delta$, so that the phase space evolution follows the Boltzmann equation. On the basis of our discussion in Section \ref{magnetizationinfinitesystem}, we expect the initial magnetization to relax to equilibrium over the timescale $\alpha^{-1}$, according to
\be
\hspace{-2cm}m(t) \sim \fr{1}{\sqrt{N}}e^{\alpha t} \mathrm{~~~~for~~~} \alpha^{-1} \ll t \ll N^\delta.
\l{mexpalpha}
\ee
Here, the prefactor accounts for fluctuations in the initial state. It follows that the magnetization acquires a value of $O(1)$ over the relaxation timescale $\tau_\mathrm{S} \sim \ln N/\alpha$, where the subscript signifies that the relaxation is due to the stochastic inter-particle collisions. 

Let us now consider the opposite limit $\alpha^{-1} \gg N^\delta$ and $\alpha^{-1} \gg t \gg N^\delta$. In this case, stochastic collisions are rare and the system evolves mostly by the deterministic dynamics. The water-bag initial state evolves not due to collisions which occur on a much longer timescale, but due to finite-size effects which act over the timescale $\sim N^\delta$. Here, similar to the result for the Vlasov-stable regime in Eq. (\ref{mstable}), the behavior of the magnetization at late times may be represented as
\be
\hspace{-2cm}m(t) \sim \fr{1}{\sqrt{N}}e^{t/N^\delta}\mathrm{~~~~for~~~}\alpha^{-1} \gg t \gg N^\delta.
\l{mexpNdelta}
\ee
In this case, the relaxation timescale $\tau_\mathrm{D}$ scales as $N^\delta \ln N$, where the subscript signifies that the relaxation to equilibrium is due to the deterministic Hamiltonian evolution.

Assuming the relaxation processes in the above two limits to be uncorrelated, the relaxation time $\tau(\alpha,N)$ may be estimated by interpolating between the two limits so that $\tau^{-1}=\tau_\mathrm{S}^{-1}+\tau_\mathrm{D}^{-1}$, thereby yielding 
\be
\hspace{-2cm}\tau(\alpha, N) \sim \fr{\ln N}{\alpha + 1/N^\delta}.
\l{T}
\ee
Equation (\ref{T}) suggests a more general scaling form:
\be
\hspace{-2cm}\tau(\alpha, N) \sim \fr{\ln N}{\alpha}s(\alpha N^{\delta}),
\l{scaling}
\ee
where, consistent with Eqs. (\ref{mexpalpha}) and (\ref{mexpNdelta}), the scaling function $s(x)$ grows as $x$ for $x \ll 1$, while $s(x) \rightarrow$ constant for $x \gg 1$. 

We now remark on the implication of Eq. (\ref{scaling}). From the scaling form, it follows that for fixed $\alpha$, the relaxation time of the water-bag initial state exhibits a crossover behavior as a function of the system size. While the relaxation time is of order $N^\delta \ln N$ (corresponding to QSS) for $N^\delta \ll \alpha^{-1}$, it becomes of order $\ln N$ for $N^\delta \gg \alpha^{-1}$. Therefore, in the presence of collisions, relaxation at long times does not occur over an algebraically growing timescale, but instead over a logarithmic timescale. This implies that under stochastic microcanonical evolution, QSS occur only as a crossover phenomenon. This scenario is similar to that observed in the canonical set-up of Ref. \cite{Baldovin:2009}. Within the canonical dynamics, however, an additional crossover takes place from the microcanonical equilibrium state to the canonical Gibbs-Boltzmann equilibrium state. 

\subsubsection{Numerical simulations}
\l{numericalsimulations}

In order to verify the physical picture put forward in Sections \ref{magnetizationinfinitesystem} and \ref{magnetizationfinitesystem}  for the behavior of magnetization and the ensuing scaling form for the relaxation time in Eq. (\ref{scaling}), we performed extensive numerical simulations of our model. The Hamilton equations (\ref{hameq}) were integrated using a symplectic fourth-order integrator with time step $dt=0.1$. In each of the $N^3$ collisions constituting an instantaneous sweep of the system, momenta of three randomly chosen particles are stochastically updated, $(p,p',p'') \rightarrow (q,q',q'')$, according to Eqs. (\ref{qdefinition}), (\ref{q'definition}) and (\ref{q''definition}) by choosing the angle $\psi$ uniformly in $[0,2\pi]$.

\bef
\begin{center}
\includegraphics[width=100mm]{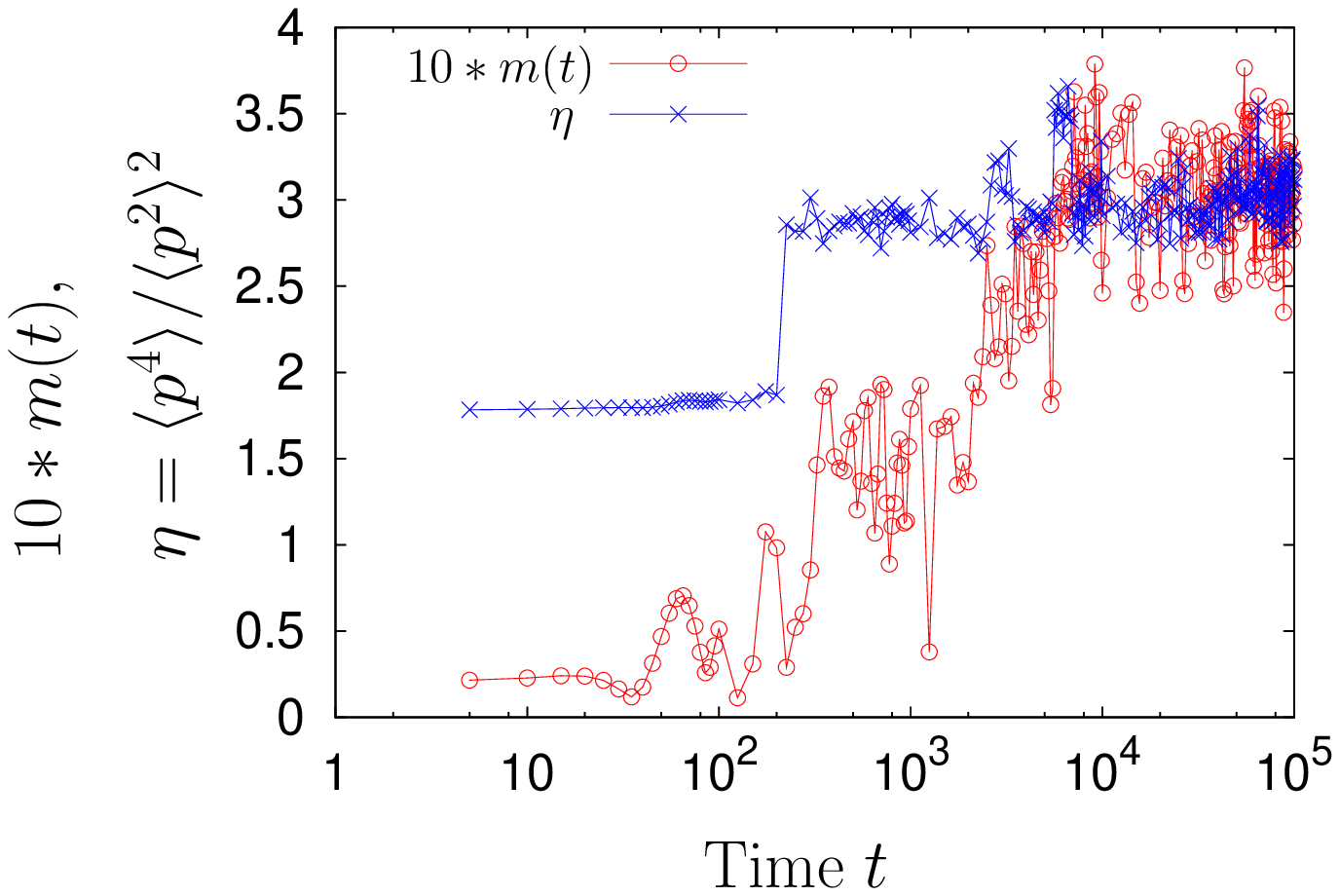}
\vspace{-5mm}
\caption{Magnetization (scaled up by a factor $10$ for convenience) and $\eta=\langle p^4 \rangle / \langle p^2 \rangle^2$ as a function of time as observed in numerical simulation of the generalized HMF model with three-particle collisions for one realization of the dynamics. Here, $N=1000$, $\eps=0.69$ and $\alpha=10^{-3}$. Note that $\alpha^{-1} \ll N^\delta$, where $\delta \simeq 1.7$ at the given energy density. Consistent with our physical arguments in Sections \ref{magnetizationinfinitesystem} and \ref{magnetizationfinitesystem}, one can observe the initial magnetization to start evolving towards equilibrium over times during which the momentum distribution becomes close to Gaussian. The latter fact is indicated by the quantity $\eta$ assuming the expected value $3$ for a Gaussian distribution.}
\l{eta-mag}
\end{center}
\eef

Figure \ref{eta-mag} shows the behavior of magnetization (scaled up by a factor $10$ for convenience) in time in our model for one realization of the dynamics while starting from a water-bag initial state. The figure also shows the value of $\eta=\langle p^4 \rangle/\langle p^2 \rangle^2$ as a function of time for the same realization of the dynamics. The angular brackets in $\eta$ represent an average over all the particles in the system. When $p$ has a Gaussian distribution, $\eta$ equals $3$. The value of $\alpha$ and the system size $N$ in Fig. \ref{eta-mag} are chosen so that the condition $\alpha^{-1} \ll N^\delta$ is satisfied. Then, on the basis of arguments in Sections \ref{magnetizationinfinitesystem} and \ref{magnetizationfinitesystem}, we expect the initial magnetization to start evolving towards equilibrium over times during which the momentum distribution becomes close to Gaussian. This can be clearly seen in Fig. \ref{eta-mag}.   

\bef
\begin{center}
\includegraphics[width=100mm]{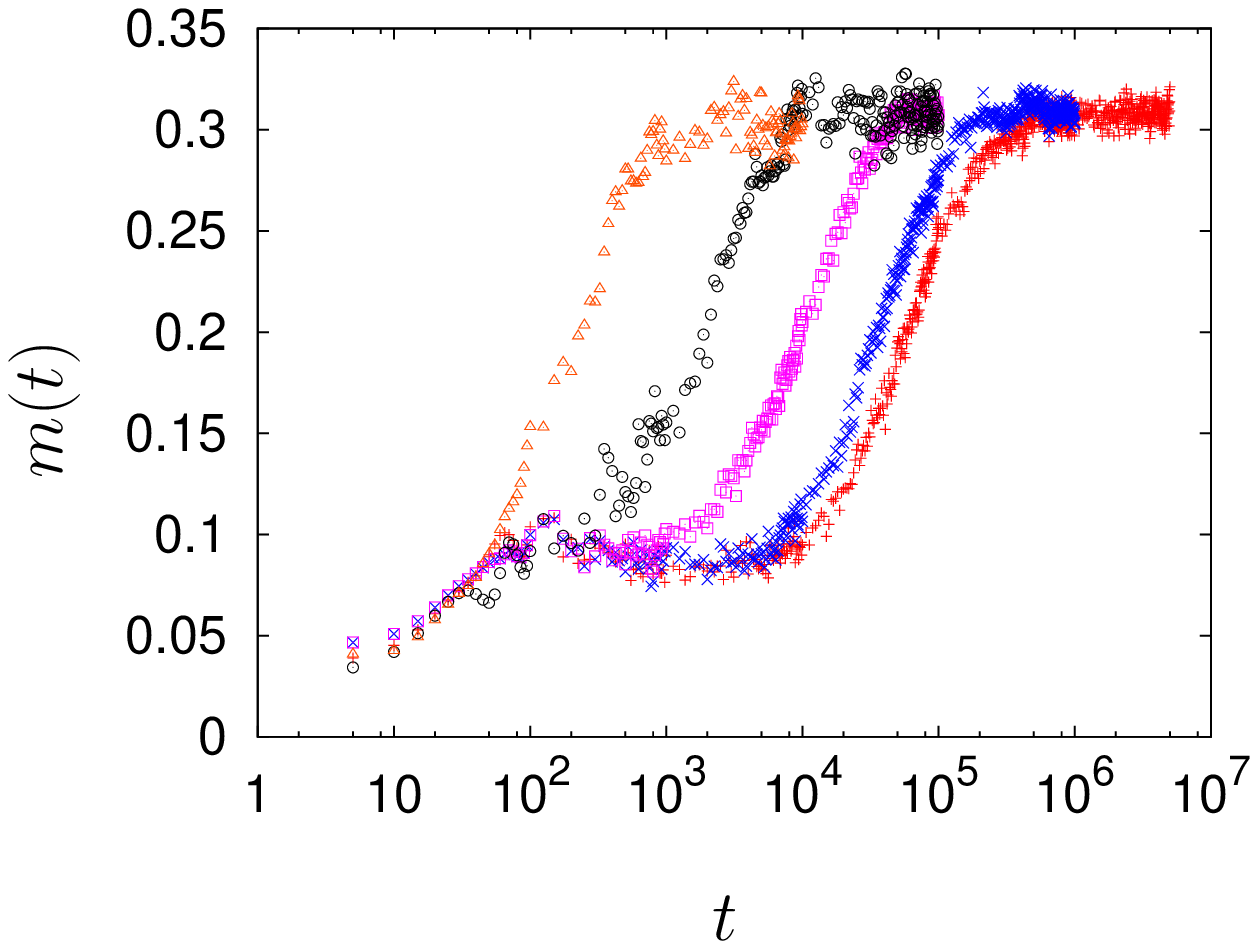}
\vspace{-5mm}
\caption{Magnetization as a function time, as observed in numerical simulations of the generalized HMF model with three-particle collisions. Here, $N=500$, energy density $\eps=0.69$ and $\alpha$ values given from right to left by $10^{-6}, 10^{-5}, 10^{-4},10^{-3},10^{-2}$. Data averaging were performed over typically hundred histories. With increasing $\alpha$, the magnetization can be seen to relax faster towards equilibrium.}
\l{3p-500}
\end{center}
\eef

Next, typical time evolutions of the magnetization for $N=500$ and several values of $\alpha$ at energy density $\eps=0.69$ are shown in Fig. \ref{3p-500}. The relaxation time $\tau(\alpha,N)$ is estimated to be the time the magnetization takes to reach the fraction $0.8$ of the final equilibrium value. Any other choice of this fraction is possible and it does not significantly affect the result. At $\eps=0.69$, when the magnetization has the equilibrium value $\simeq 0.3$ and $\delta \simeq 1.7$ \cite{Yamaguchi:2004}, plotting $\alpha \tau(\alpha, N)/\ln N$ against $\alpha N^\delta$ shows a very good scaling collapse over several decades, as shown in Fig. \ref{fig2}. This is in accordance with our scaling form, Eq. (\ref{scaling}), for the relaxation time and supports our prediction for QSS as a crossover phenomenon under noisy dynamics within a microcanonical ensemble.

\bef
\begin{center}
\includegraphics[width=100mm]{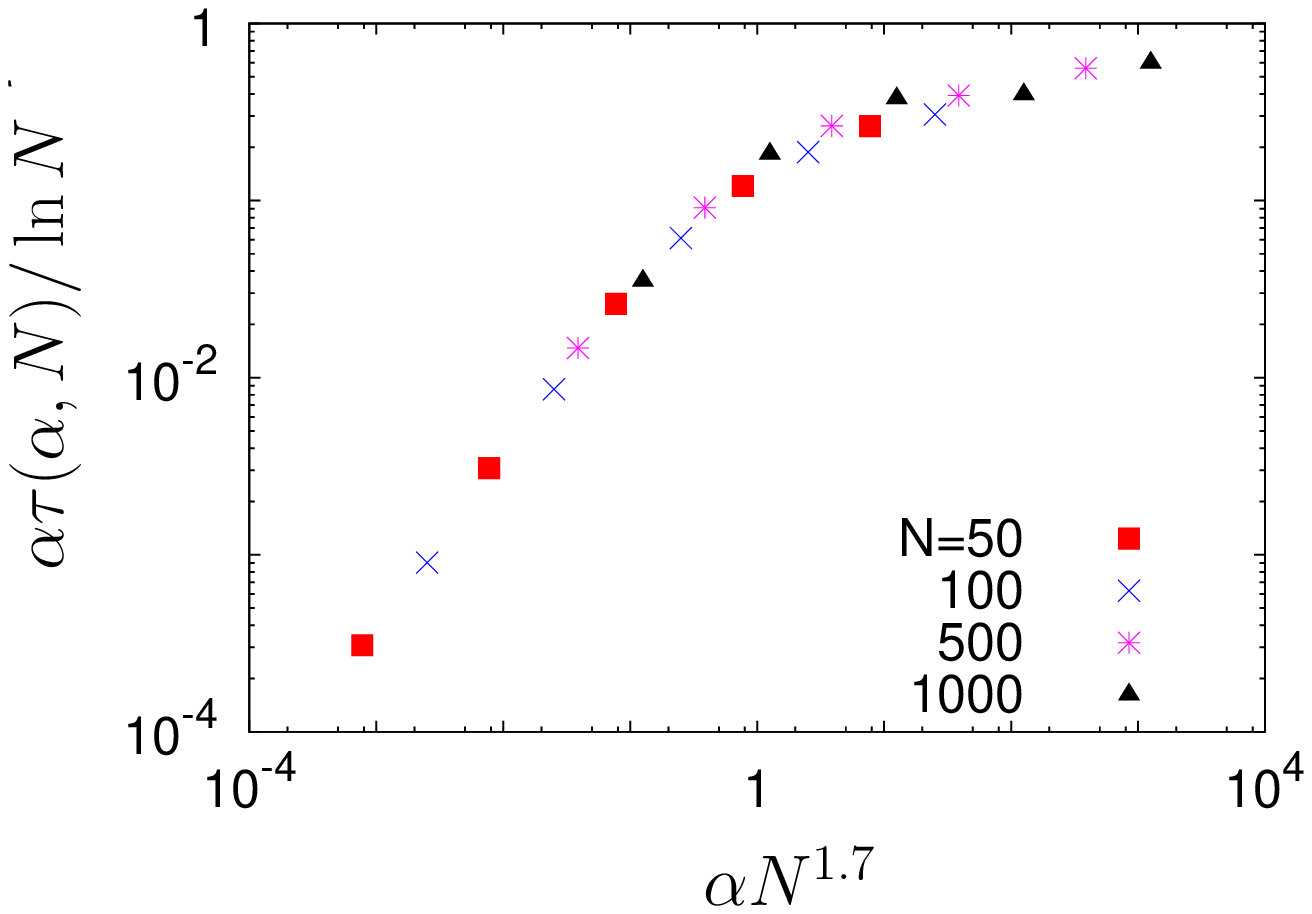}
\vspace{-5mm}
\caption{$\alpha \tau(\alpha, N)/\ln N$ vs. $\alpha N^\delta$, based on the data obtained from numerical simulations of the generalized HMF model with three-particle collisions. Here, $\eps=0.69$ for which $\delta \simeq 1.7$ \cite{Yamaguchi:2003, Yamaguchi:2004}. The system sizes are marked in the figure. Data averaging varies between $5 \times 10^4$ histories for the smallest system and $100$ histories for the largest one. One observes a very good scaling collapse, thereby supporting Eq. (\ref{scaling}).}
\l{fig2}
\end{center}
\eef

\section{Generalized HMF model with microcanonical Monte Carlo stochastic dynamics}
\l{2collisions}
In order to test the validity of our prediction for QSS in a different framework for the noisy dynamics, we next study our generalized HMF model with the stochastic process of two-particle collisions discussed in Section \ref{introduction}. We then investigate numerically our prediction for QSS as a crossover phenomenon through verifying the scaling form, Eq. (\ref{scaling}), for the relaxation time to equilibrium.
\subsection{The model}
\l{themodelCreutz}
In our model, the stochastic process of two-particle collisions is carried out according to the algorithm for microcanonical Monte Carlo simulation developed by Creutz \cite{Creutz:1983}. In the spirit of this algorithm, we introduce an extra degree of freedom, called the demon, with initial energy $E_D=0$. The system energy $E_S$ has the given initial value $E$. During the dynamics of the model, the combined total energy of the system and the demon, $E_S+E_D$, remains conserved. Unlike the generalized HMF model with three-particle collisions in Section \ref{themodel}, the energy of the system during the dynamics fluctuates and all microscopic configurations of the system with energy $E_S \le E$ are sampled with equal probability \cite{Creutz:1983}. 

Similar to the model in Section \ref{themodel}, here the model evolves according to the following repetitive sequence of events. A deterministic evolution according to Eq. (\ref{hameq}) for a random time interval $\tau$ is followed by an instantaneous sweep which consists of $N^2$ collisions. The time interval $\tau$ is distributed as $\alpha e^{-\alpha \tau}$. Each collision involves the following steps:
\begin{enumerate}
\item{Choose a pair of random spins $i$ and $j$ with momentum $p_i$ and $p_j$, respectively.
\l{step2}}
\item{Since total momentum is conserved in collisions, we attempt to update the momenta $p_i, p_j$ in the following way:
\bea
p_i \rightarrow q_i=p_i+\xi \nonumber, \\
\l{pupdate} \\
p_j \rightarrow q_j=p_j-\xi \nonumber.
\eea
Here, $\xi$ is a random number uniformly distributed in an arbitrary interval symmetric about zero. It can be easily seen that such a distribution for $\xi$ ensures that all configurations of the system with energy $E_S \le E$ are sampled with equal probability. 
}
\item{Next, compute the change $\Delta E_S$ in the energy of the system due to the attempted momentum update: $\Delta E_S=\xi(\xi+p_i-p_j)$.}
\item{For $\Delta E_S <0$ or for $0 < \Delta E_S < E_D$, the momentum update is implemented and the demon energy is updated according to $E_D \rightarrow E_D - \Delta E_S$. Otherwise, the update is rejected. Thus, the momentum updates, Eq. (\ref{pupdate}), are actually carried out only if the demon possesses the required amount of energy for the update. This completes one collision.}
\end{enumerate}

In the limit of large system size, the demon energy represents only a small fraction of the total energy. Hence, during the dynamics, the system energy fluctuates within a finite energy band, and we expect the prediction of Section \ref{stochastic} for the existence of QSS as a crossover phenomenon to also hold in the present case. In particular, we now proceed to verify the scaling form, Eq. (\ref{scaling}), in numerical simulations.  
\subsection{Numerical results}
\bef
\begin{center}
\includegraphics[width=100mm]{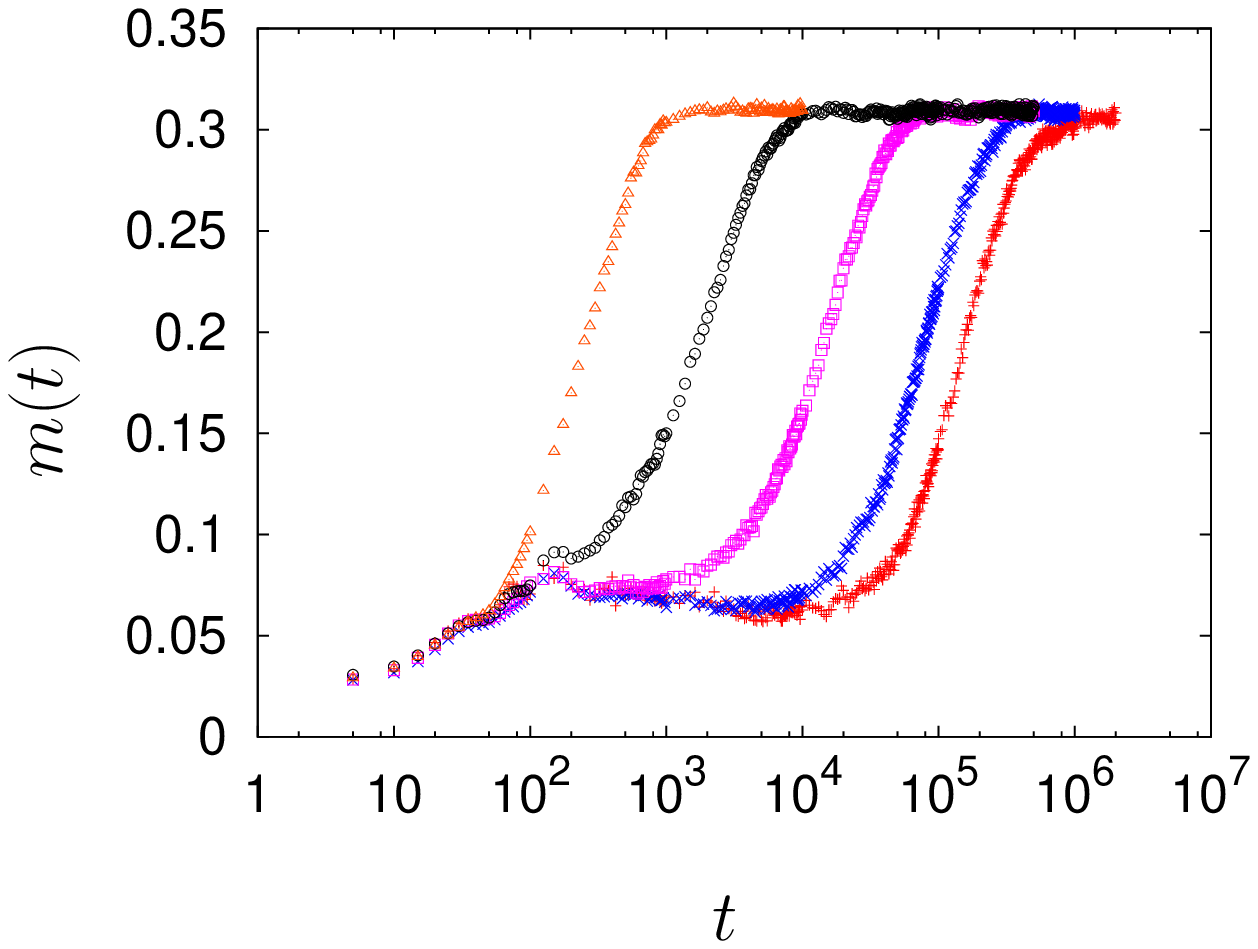}
\vspace{-5mm}
\caption{Magnetization as a function time, on the basis of numerical simulations of the generalized HMF model with two-particle collisions. Here, $N=1000$, $\eps=0.69$ and $\alpha$ values given from right to left by $10^{-6}, 10^{-5}, 10^{-4},10^{-3}$ and $10^{-2}$. Data averaging were performed over several hundred histories. With increasing $\alpha$, the magnetization is seen to relax faster towards equilibrium. The behavior of the magnetization is similar to that in Fig. \ref{fig2}.}
\l{2p-1000}
\end{center}
\eef
Here, we present results from our numerical simulations for the behavior of magnetization as a function of time while starting from a water-bag state. The Hamilton equations of motion were integrated using a symplectic fourth-order integrator with time step $dt= 0.1$. Typical time evolutions of the magnetization for a system of size $N=1000$ and several values of $\alpha$ are shown in Fig. \ref{2p-1000} for the energy density $\eps=0.69$. The behavior of magnetization is similar to that in Fig. \ref{fig2} for the generalized model with three-particle collisions. 

\bef
\begin{center}
\includegraphics[width=100mm]{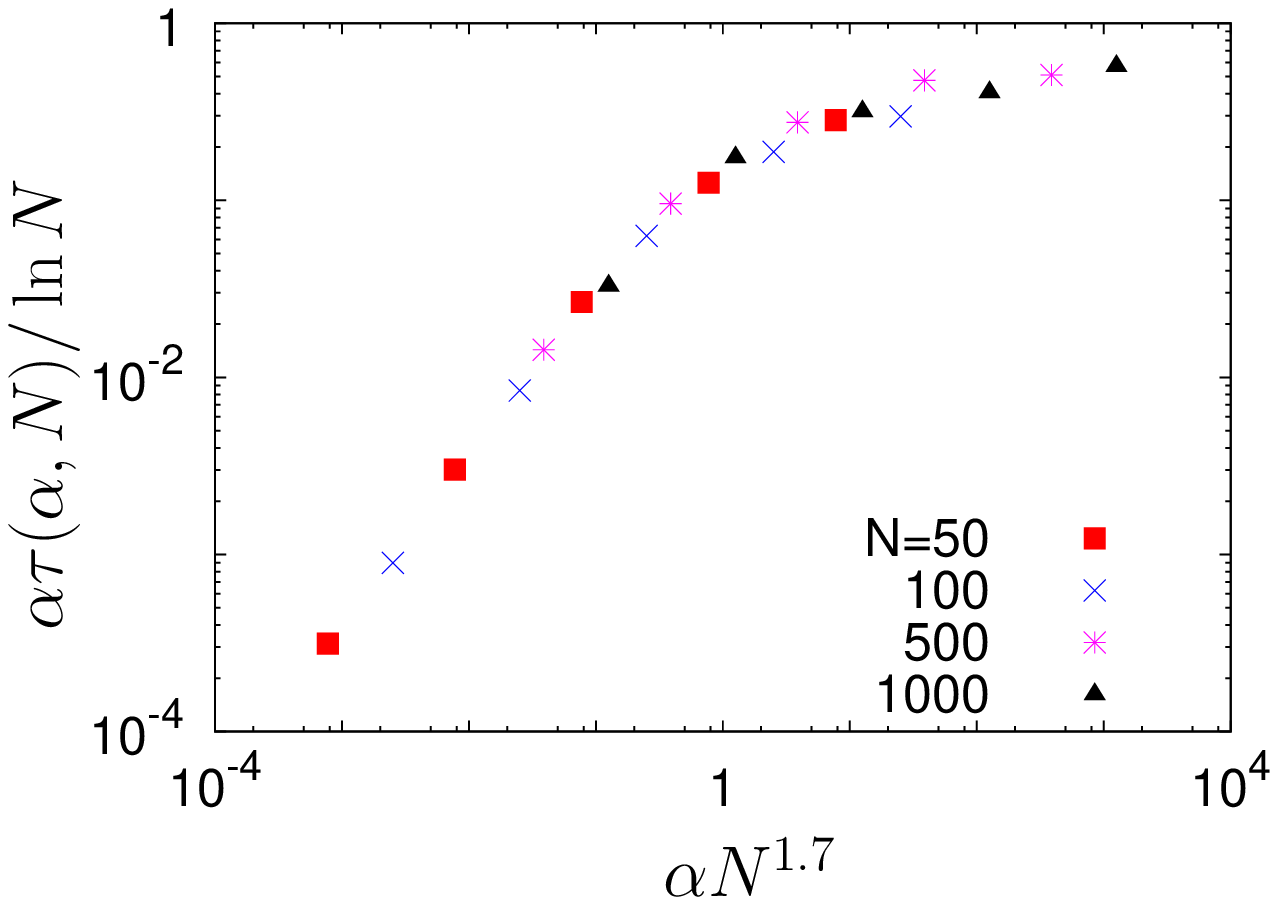}
\vspace{-5mm}
\caption{$\alpha \tau(\alpha, N)/\ln N$ vs. $\alpha N^\delta$, based on the data obtained from numerical simulations of the generalized HMF model with two-particle collisions. Here, $\eps=0.69$, $\delta \simeq 1.7$, the system sizes are marked in the figure. Data averaging varies between $5 \times 10^4$ histories for the smallest system and a few hundred histories for the largest one. A very good scaling collapse in accordance with Eq. (\ref{scaling}) may be seen.}
\l{fig4}
\end{center}
\eef

We measure the relaxation time $\tau(\alpha,N)$ by a method similar to that discussed in Section \ref{numericalsimulations}. At energy density $\eps=0.69$ for which $\delta \simeq 1.7$, plotting $\alpha \tau(\alpha,N)/\ln N$ as a function of $\alpha N^\delta$ gives a very good scaling collapse in accordance with Eq. (\ref{scaling}), as shown in Fig. \ref{fig4}. In addition to the results in Section \ref{stochastic}, this further supports Eq. (\ref{scaling}) and our prediction for QSS as a crossover phenomenon under stochastic dynamics within a microcanonical ensemble. 

\section{Concluding remarks}
\l{conclusions}
In this paper, we addressed the robustness of quasistationary states (QSS) in long-range interacting systems with respect to non-deterministic dynamics within a microcanonical ensemble. We considered a paradigmatic long-range interacting system, the Hamiltonian Mean-Field model, which is known to exhibit QSS under deterministic dynamics. We generalized the model to include stochastic dynamical moves in addition to the deterministic ones. Our model evolves by a piecewise deterministic dynamics whereby deterministic Hamiltonian evolution is randomly interrupted by stochastic dynamical processes. We considered two different stochastic processes, namely, (i) three-particle collisions, and (ii) two-particle collisions. Both lead to stochastic updates of momenta of the colliding particles while conserving the total momentum of the system. In (i), the energy of the system is strictly conserved in collisions. In (ii), however, the energy is not conserved during collisions, instead fluctuates below the given initial value within a finite energy band, which is negligible in the thermodynamic limit. Our analysis suggests that within the ambit of our generalized model, QSS occur only as a crossover phenomenon over times which are determined by the strength of the stochastic process. In particular, we showed that in the limit of long times, there are no QSS under stochastic dynamics within a microcanonical ensemble. We proposed a scaling form for the relaxation time to equilibrium and verified it by extensive numerical simulations. It would be of interest to investigate the general validity of our results in other models with long-range interactions.            

\section{Acknowledgments}
We thank A. C. Barato, O. Cohen, T. Dauxois, O. Hirschberg, H. Posch and S. Ruffo for fruitful discussions and useful comments on the manuscript. We gratefully acknowledge the support of the Israel Science Foundation (ISF) and the Minerva Foundation with funding from the Federal German Ministry for Education and Research.
\appendix
\section{Evaluation of the integral $I$ in Eq. (\ref{Idefinition})}
\l{appendix1}
In order to evaluate the integral,
\be
\hspace{-2cm}I=\int dp dp'dp''dqdq'dq'' ~f_0(p)f_0(p')f_0(p'')R\phi^{0}_{\pm 1}(q,\o_0),
\l{Idefinitionapp}
\ee
one expresses the updated momenta $(q,q',q'')$ as in Eqs. (\ref{qdefinition}), (\ref{q'definition}) and (\ref{q''definition}), so that one has \cite{Ma:1983}
\be
\hspace{-2cm}dpdp'dp''dqdq'dq''R=\fr{\alpha}{3}r d\psi d\psi' dP dr.
\l{integrationmeasuredefinition2}
\ee
Here, as explained in the paragraph following Eq. (\ref{nudefinition}), $P$ and $E$ are the three-particle momentum and energy, given respectively by Eq. (\ref{Pdefinition}) and Eq. (\ref{Edefinition}). The quantity $r=\sqrt{2E-P^2/3}$ gives the radius of the circle formed by the intersection of the plane given by Eq. (\ref{Pdefinition}) and the spherical surface given by Eq. (\ref{Edefinition}). Note that $P \in [-\infty,\infty]$, while $r \in [0, \infty]$. Moreover, $\psi, \psi' \in [0,2\pi]$ are angles measured along the circle of intersection. 

We have, from Eq. (\ref{fp}),
\be
\hspace{-2cm}f_0(p)f_0(p')f_0(p'')=\fr{1}{(2\pi T)^{3/2}}e^{-(r^2+P^2/3)/2T}.
\l{fproduct}
\ee

We use Eqs. (\ref{unperturbedfn}), (\ref{integrationmeasuredefinition2}) and (\ref{fproduct}) in Eq. (\ref{Idefinitionapp}). We then scale the variables $P,r,q$ by the temperature $T$. Finally, we use Eq. (\ref{qdefinition}) to get
\be
\hspace{-2cm}I=\fr{\alpha I_\pm}{6T(2\pi)^{3/2}}\int r d\psi d\psi' dP dr ~e^{-(r^2+P^2/3)/2}\fr{\fr{P}{\sqrt{3}}+\sqrt{\fr{2}{3}}r\cos \psi}{\fr{P}{\sqrt{3}}+\sqrt{\fr{2}{3}}r\cos \psi \pm \o_0/\sqrt{T}}. 
\ee
The above equation can be rewritten as
\bea
&&\hspace{-2cm}I=\fr{\alpha 4\pi^2 I_\pm}{6T(2\pi)^{3/2}}\int r dP dr ~e^{-(r^2+P^2/3)/2}\mp\fr{\o_0\alpha 2\pi I_\pm}{6(2\pi T)^{3/2}}\int \fr{r d\psi dP dr ~e^{-(r^2+P^2/3)/2}}{\fr{P}{\sqrt{3}}+\sqrt{\fr{2}{3}}r\cos \psi\pm\o_0}\nonumber \\
&&\hspace{-1.8cm}=\fr{\alpha \pi I_\pm}{T\sqrt{3}}\mp\fr{\o_0\alpha \pi I_\pm}{3(2\pi T)^{3/2}}\int_{-\pi}^{\pi} d\psi \int_{-\infty}^{\infty}dP \int_0^{\infty} \fr{r dr ~e^{-(r^2+P^2/3)/2}}{\fr{P}{\sqrt{3}}+\sqrt{\fr{2}{3}}r\cos \psi\pm\o_0},
\eea
where we have used the result $\int_{-\infty}^\infty dP \int_0^\infty rdr ~e^{-(r^2+P^2/3)/2}=\sqrt{6\pi}$. Manipulating the second integral on the right so that both $P$ and $\cos \psi$ assume only positive values, we obtain
\bea
&&\hspace{-2cm}I=\fr{\alpha \pi I_\pm}{T\sqrt{3}}+
\fr{\o_0^2\alpha I_\pm}{3T^{3/2}\sqrt{2\pi}}\int_{-\pi/2}^{\pi/2} d\psi \int_0^{\infty}dP \int_0^{\infty}r dr ~e^{-\fr{1}{2}(r^2+P^2/3)} \nonumber \\
&&\times \left(\fr{1}{\left(\fr{P}{\sqrt{3}}-\sqrt{\fr{2}{3}}r\cos \psi\right)^2-\o_0^2}+\fr{1}{\left(\fr{P}{\sqrt{3}}+\sqrt{\fr{2}{3}}r\cos \psi\right)^2-\o_0^2}\right).
\l{Imiddlestep}
\eea
We recall that the unperturbed frequency $\o_0$ obeys $\o_0^2=-\O_0^2$, with real $\O_0 \ge 0$. Substituting $\o_0^2=-\O_0^2$ and also using the fact that the critical temperature $T_c=1/2$ in Eq. (\ref{Imiddlestep}), we arrive at the following result.
\be
\hspace{-2cm}I=\fr{2\pi \alpha T_c I_\pm}{T\sqrt{3}}-\fr{\O_0^2\alpha I_\pm}{3T^{3/2}\sqrt{2\pi}}\left(I_1+I_2\right),
\l{Ifinal}
\ee
where
\be
\hspace{-2cm}I_1=\int_{-\pi/2}^{\pi/2} d\psi \int_0^{\infty}dP \int_0^{\infty}\fr{r dr ~e^{-\fr{1}{2}(r^2+P^2/3)}}{\left(\fr{P}{\sqrt{3}}-\sqrt{\fr{2}{3}}r\cos \psi\right)^2+\O_0^2}
\l{I1definition}
\ee
and
\be
\hspace{-2cm}I_2=\int_{-\pi/2}^{\pi/2} d\psi \int_0^{\infty}dP \int_0^{\infty}\fr{r dr ~e^{-\fr{1}{2}(r^2+P^2/3)}}{\left(\fr{P}{\sqrt{3}}+\sqrt{\fr{2}{3}}r\cos \psi\right)^2+\O_0^2}.
\l{I2definition}
\ee

Now, we will evaluate the integrals $I_1$ and $I_2$ to leading order in $\O_0$. It will turn out that, to leading order, only $I_1$ contributes to $I$. To compute $I_1$, we make the substitution $y=\fr{P}{\sqrt{3}}-\sqrt{\fr{2}{3}}r\cos \psi$ to rewrite $I_1$ as
\be
\hspace{-2cm}I_1=\int_{-\pi/2}^{\pi/2} d\psi \int_0^{\infty}re^{-\fr{r^2}{2}(1+2(\cos^2\psi)/3)} dr \int_{-\sqrt{\fr{2}{3}}\cos \psi}^\infty dy ~\fr{e^{-\fr{1}{2}(y^2+2\sqrt{\fr{2}{3}}ry\cos \psi)}}{y^2+\O_0^2}.
\ee
Next, with the substitution $z=y/\O_0$, we obtain
\be
\hspace{-2cm}I_1=\fr{1}{\O_0}\int_{-\pi/2}^{\pi/2} d\psi \int_0^{\infty}re^{-\fr{r^2}{2}(1+2(\cos^2\psi)/3)} dr \int_{-\sqrt{\fr{2}{3}}\fr{\cos \psi}{\O_0}}^\infty dz ~\fr{e^{-\fr{1}{2}\left(z^2\O_0^2+2\sqrt{\fr{2}{3}}rz\O_0\cos \psi \right)}}{z^2+1}.
\ee
To leading order in $\O_0$, we get
\be
\hspace{-2cm}I_1=\fr{1}{\O_0}\int_{-\pi/2}^{\pi/2} d\psi \int_0^{\infty}re^{-\fr{r^2}{2}(1+2(\cos^2\psi)/3)} dr \left[\pi+O(\O_0)\right],
\l{I1final}
\ee
where we have used
\be
\hspace{-2cm}\int_{-\infty}^\infty \fr{dz}{z^2+1}=\pi.
\ee
Following similar steps, we obtain
\be
\hspace{-2cm}I_2=\fr{1}{\O_0}\int_{-\pi/2}^{\pi/2} d\psi \int_0^{\infty}re^{-\fr{r^2}{2}(1+2(\cos^2\psi)/3)} dr \left[0+O(\O_0)\right],
\l{I2final}
\ee
It therefore follows on substituting in Eq. (\ref{Ifinal}) that, to leading order in $\O_0$, $I_2$ does not contribute to $I$. 

Finally, using Eqs. (\ref{I1final}) and (\ref{I2final}) in Eq. (\ref{Ifinal}), we get
\bea
&&\hspace{-2cm}I=\fr{2\pi \alpha T_c I_\pm}{T\sqrt{3}}-\fr{\O_0 \alpha I_\pm}{3T^{3/2}\sqrt{2\pi}}\int_{-\pi/2}^{\pi/2} d\psi \int_0^{\infty}re^{-\fr{r^2}{2}(1+2(\cos^2\psi)/3)} dr \left[\pi+O(\O_0)\right] \nonumber \\
&&\hspace{-1.8cm}=\fr{2\pi \alpha T_c I_\pm}{T\sqrt{3}}-\fr{\pi^{3/2}\alpha I_\pm}{T^{3/2}\sqrt{30}}\left[\O_0 +O(\O_0^2)\right],
\l{Ifinalresult}
\eea
where, in obtaining the last line, we have used $\int_{-\pi/2}^{\pi/2} d\psi \int_0^{\infty}re^{-\fr{r^2}{2}(1+2(\cos^2\psi)/3)} dr=\sqrt{\fr{3}{5}}\pi$. 
\section*{References}

\end{document}